\documentstyle[12pt]{article}
\begin{document}
\large

\newpage
\begin{center}
{\bf MASS OF THE NEUTRINO AND ITS CHARGE}
\end{center}
\vspace{1cm}
\begin{center}
{\bf Rasulkhozha S. Sharafiddinov}
\end{center}
\vspace{1cm}
\begin{center}
{\bf Institute of Nuclear Physics, Uzbekistan Academy of Sciences,
Tashkent, 702132 Ulugbek, Uzbekistan}
\end{center}
\vspace{1cm}

Both electron and its neutrino possess not only the anomalous magnetic
moment but each of existing types of the electric charges and their
dipole moments. Any of them can interact with field of emission leading
to the elastic scattering on a spinless nucleus. We list some implications
implied from the analysis of these phenomena in the case of one - photon
exchange. The processes cross sections give the possibility to establish
the individual as well as the united equation between the Dirac and the
Pauli form factors of light leptons. They define the electronic neutrino
electric charge as a consequence of the availability of a kind of mass.

\newpage
In many works a question about the neutrino with a non - zero rest mass
was investigated. Analysis of existing experiments assumed \cite{1} that
a massive Dirac neutrino must have not only the magnetic moment \cite{2}
but also the electric charge. Such a conclusion one can make by following
the behavior of the neutrino in the nucleus field. From this point of view,
the spin phenomena \cite{3} may become highly useful.

In a given work the elastic scattering of electrons and their neutrinos
on a spinless nucleus have been considered at the account of fermions charge
and magnetic moment interactions with field of emission of virtual photons.
Starting from the processes cross sections equality for longitudinal polarized
and unpolarized particles, it is shown that if the neutrino corresponds to
the electron $(\nu=\nu_{e}),$ between the Dirac and the Pauli form factors of
the neutrino and electron there exists the individual as well as the united
dependence. They state that the four - component neutrino possesses both
normal and anomalous electric charges. Herewith its full electric charge
has the size
\begin{equation}
e_{\nu}=-\frac{3eG_{F}m_{\nu}^{2}}{4\pi^{2}\sqrt{2}}, \, \, \, \, e=|e|.
\label{1}
\end{equation}

Such a regularity, however, meets with many problems which give 
the possibility to make the most diverse predictions. The treatment 
of any of them would bring us too far and all they therefore will require 
the more detailed description. But here we can add the following.

Using (\ref{1}) for the neutrino \cite{4} with mass $m_{\nu}=10\ {\rm eV}$
and taking into account \cite{5} that
$G_{F}=1.16637\cdot 10^{-5}\ {\rm GeV^{-2}},$ we find
\begin{equation}
e_{\nu}=6.267\cdot10^{-25}
\left(\frac{m_{\nu}}{1\ {\rm eV}}\right)^{2}\ {\rm e}=
6.27\cdot10^{-23}\ {\rm e}.
\label{2}
\end{equation}

Earlier laboratory facts and conservation of charge in neutron decay define
the upper limit \cite{6} equal to $e_{\nu}< 4\cdot10^{-17}\ {\rm e}.$
Refinement of each of the electric charges of the neutron, proton and
electron allows to conclude \cite{1} that $e_{\nu}< 10^{-21}\ {\rm e}.$
Analysis of elastic $\overline{\nu_{e}}e$ scattering experiment \cite{7}
assumed \cite{8} that $e_{\nu}< 2.7\cdot10^{-10}\ {\rm e}.$ Cosmological
considerations for the neutrino charge lead to the estimate \cite{9}
of $e_{\nu}< 10^{-17}\ {\rm e}.$

We recognize that (\ref{2}) violates the charge conservation law. There
are many uncertainties both in the nature and in the size of the neutrino
mass. Another reason of inconsistency is the absence of quality picture
of $\beta -$ decay processes. Nevertheless, if we suppose \cite{1} that
$e_{\nu}< 10^{-21}\ {\rm e}$ then taking (\ref{1}), it is not difficult
to establish the theoretical bound on the neutrino mass:
$m_{\nu}< 40\ {\rm eV}.$ It is compatible with that following
from the experiment \cite{10}: $14\ {\rm eV}< m_{\nu}< 46\ {\rm eV}.$

Basing on the analysis of evolution of the Universe, it was found \cite{11}
that $m_{\nu}< 0.3\ {\rm eV}.$ Insertion of this value in (\ref{1})
gives $e_{\nu}< 5.64\cdot 10^{-26}\ {\rm e}.$

Having the formula (\ref{1}) and by following the fact that the force
of the Newton attraction between the two neutrinos is less than the force
of their Coulomb repulsion, we get the following estimates of
\begin{equation}
m_{\nu}>\frac{4\pi^{2}\sqrt{2}}{3G_{F}}
\left(\frac{G_{N}}{\alpha}\right)^{1/2}=1.53\cdot 10^{-3}\ {\rm eV},
\label{3}
\end{equation}
\begin{equation}
e_{\nu}>\frac{4\pi^{2}\sqrt{2}}{3G_{F}}
\left(\frac{G_{N}}{\alpha}\right)\ {\rm e}=1.46\cdot 10^{-30}\ {\rm e},
\label{4}
\end{equation}
where $G_{N}$ is the constant of gravitational emission.

Of course, such a definition of values of (\ref{3}) and (\ref{4}) is not
very standard. At the same time the existing laboratory bounds may serve
as further confirmations of our earlier findings. Insofar as the discrepancy
is concerned, it reflects just many properties of a certain latent regularity
of general picture of massive neutrinos.

In the framework of the loop phenomena, the neutrino must be electrically
neutral \cite{12} at the condition of gauge invariance. It appears that
here on the basis of (\ref{1}) one can will decide a question about
the equality of the neutrino physical mass to zero. But we can say that
a non - zero interaction of Pauli arises at the expense of usual Dirac
interaction. Thus, the neutrality of the neutrino in the loop approximation
one must interpret as an indication to the new structure of electromagnetic
gauge invariance.

\end{document}